\documentclass{pramana}
\usepackage{microtype}
\usepackage{amsmath}
\usepackage{amsfonts}
\usepackage{graphicx}  
\usepackage{cite}
\usepackage{hyperref}
\usepackage{xcolor}
\usepackage{booktabs}
\usepackage{float}

\begin{document}

\title{Generation and Enhancement of Bipartite and Tripartite Entanglement in an Electro-Optomechanical Ring Cavity}


\author{Fouad Essaadi\textsuperscript{1}, Yassine Oussarhan\textsuperscript{1}, Mohamed Ouhammou\textsuperscript{2}, 
	Said Mouslih\textsuperscript{1}, Mohamed Jakha\textsuperscript{1}, Bouzid Manaut\textsuperscript{1,*}\and 
	Souad Taj\textsuperscript{1}}
\affilOne{\textsuperscript{1} Laboratoire de Recherche Pluridisciplinaire en Physique (L.R.P.P), Polydisciplinary Faculty, Sultan Moulay Slimane University, Beni Mellal, 23000, Morocco.\\}
\affilTwo{\textsuperscript{2} Ecole Supérieure de l’Education et de la Formation, Sultan Moulay Slimane University, Beni Mellal, 23000, Morocco.}


\twocolumn[{

\maketitle

\corres{b.manaut@usms.ma}


\begin{abstract}
This study investigates the generation and enhancement of quantum entanglement in an electro-optomechanical ring cavity system. The setup integrates two Coulomb-coupled mechanical resonators, which serve as the fundamental mechanism for the generation of bipartite and tripartite entanglement via charge mediated coupling. We then demonstrate the significant enhancement of this entanglement via a nonlinear parametric drive (an optical parametric amplifier, OPA), which injects a controllable nonlinearity into the cavity. We derive the system's Hamiltonian and the corresponding quantum Langevin equations, which are linearized around steady-state solutions to analyze Gaussian quantum fluctuations. Employing the covariance matrix formalism, we quantify bipartite entanglement via logarithmic negativity and tripartite entanglement via the minimum residual contangle. Our results unequivocally show that while the Coulomb interaction is indispensable for creating entanglement, the OPA acts as a powerful control tool, dramatically amplifying the degree of quantum correlations for all subsystems. We find that the strength of entanglement is highly sensitive to several parameters and can be optimized through the strategic selection of the OPA's gain and phase, the laser detuning, and the input power. A key finding is the existence of a trade-off, where parameters that maximize entanglement also constrain the stable operating regime of the system. Furthermore, thermal noise is shown to progressively degrade all quantum correlations, underscoring the necessity for low-temperature operation. These findings provide comprehensive guidance for parameter optimization, outlining a clear path from generation to enhancement, and highlight the potential of such hybrid systems as versatile platforms for controlling multipartite entanglement in quantum technologies.
\end{abstract}
}] 

\doinum{12.3456/s78910-011-012-3}
\artcitid{\#\#\#\#}
\volnum{123}
\year{2026}
\pgrange{1--12}
\setcounter{page}{1}
\lp{12}

\section{Introduction}

The story of quantum mechanics is one of revolutionary ideas and unexpected discoveries, beginning with the foundational debates of the early 20th century \cite{Jammer1966,Pais1982}. Among these, the phenomenon of quantum entanglement stands out as a concept that has both challenged our understanding of nature and opened new frontiers in science and technology\cite{Aspect2002}. First brought into the spotlight by Albert Einstein and his colleagues Podolsky and Rosen when they challenged quantum mechanics through their seminal 1935 paper \cite{einstein1935can} and has since emerged as an essential element of modern physics\cite{Bell1987}.

Entanglement has been confirmed experimentally through violations of Bell's inequalities \cite{bell1964on,aspect1981experimental,aspect1982experimental,aspect1982experimental2}. The nonlocal correlations between spatially separated quantum systems are a vital component in quantum information \cite{Bennett1984,Nielsen2010}.
 Over the past few decades, entanglement has become a fundamental resource in quantum technologies, enabling advancements in quantum communication \cite{kimble2008quantum}, quantum cryptography \cite{ekert1991quantum}, and quantum computing \cite{pathak2013elements}. The experimental realization and manipulation of this phenomenon has driven extensive research across multiple platforms including photonic systems through atomic ensembles to newer hybrid optomechanical setups \cite{aspelmeyer2014,kippenberg2008}. Optomechanical systems create opportunities for quantum mechanics research and quantum technology development through optical-mechanical coupling by radiation pressure \cite{Bowen2015}. Optomechanical systems are generally made up of an optical cavity paired with a movable mechanical element that interacts with the electromagnetic field trapped inside the cavity through radiation pressure \cite{caves1980}. This pressure arises from the momentum transferred by photons as they reflect off the surface of the mechanical component such as a vibrating mirror or a tiny membrane \cite{Xia2023}. As a result, this force causes minute shifts in the position of the mechanical element, altering the cavity's length and, consequently, the properties of the confined light field. These changes become particularly noticeable when the cavity is powered by a robust laser source \cite{simon2011}.
Optomechanical systems, however, offer a unique advantage by coupling the microscopic world of photons with the macroscopic motion of mechanical objects. This force exerted by light on matter, enables researchers to probe quantum effects at scales previously thought to be governed solely by classical mechanics. Such systems have become testbeds for exploring the boundaries of quantum theory, including the decoherence of quantum states and the potential for macroscopic quantum phenomena \cite{youssefi2023}. The rising prominence of optomechanical systems stems from their capability to link macroscopic objects with quantum mechanics. Research breakthroughs in optomechanics include both achieving quantum ground state cooling of mechanical resonators \cite{chan2011} and generating optical-mechanical entanglement \cite{vitali2007,amazioug2018} through light-mechanical oscillator interactions. The field of cavity optomechanics is receiving significant attention recently due to its applications in precision measurement \cite{xu2022optomechanically,li2021cavity,brawley2016nonlinear}, quantum information processing \cite{kippenberg2007cavity,marquardt2011quantum} and fundamental tests of quantum mechanics.

Ring cavities distinguish themselves from other configurations with their distinct benefits. The closed-loop resonator design maintains circulating optical modes which strengthen light-matter interactions to create a reliable and effective system for quantum state engineering \cite{huang2009entangling,schulze2010optomechanical}. The cyclic nature of energy transfer in ring cavities amplifies interaction strengths, making them ideal for studying quantum correlations.
Optomechanical research has undergone a major advancement through the integration of nonlinear optical components including OPAs. Previous studies have demonstrated that OPAs can enhance optomechanical coolings \cite{huang2009}, prepare squ-eezed states of the mechanical mode \cite{agarwal2016}, and improve entanglement in cavity optomechanical systems \cite{mi2013,hu2017,xuereb2012}.

The control of multipartite entanglement in hybrid quantum systems operates within the broader framework of nonlinear dynamics engineering. Contemporary research leverages integrable system theory and symbolic computation to extract complex coherent structures like multi-solitons and breathers from variable-coefficient models in ferromagnetic materials, plasmas, and shallow-water wave dynamics \cite{gao2025kmm, gao2025sk, gao2025hs}. These studies, along with analyses of models in cosmic plasmas and forced fluid environments, systematically reveal how external parameters and variable coefficients acting as control knobs for nonlinearity and dispersion dictate the stability and morphology of nonlinear waves \cite{gao2025zkb}. Transposing this paradigm to the quantum regime, our electro-optomechanical cavity integrates two fundamental nonlinear controls: the Coulomb coupling and the parametric nonlinearity of an OPA. We treat the OPA's parameters as precisely tunable synthetic coefficients, analogous to those in classical wave systems, to dynamically sculpt the Hamiltonian for enhanced entanglement \cite{gao2024water, shan2024kdv, feng2025fluid}. Finally, the resulting parameter-dependent entanglement landscapes and stability trade-offs constitute a quantum analog to nonlinear wave management, establishing a unified perspective on harnessing engineered nonlinearities across physical scales \cite{wang2026hd, gao2025kvcbs}.

Entanglement generation in electro-optomechanical systems has recently been the focus of numerous theoretical proposals \cite{zhang2012,bai2017}. A particularly promising architecture utilizes Coulomb-interacting mechanical oscillators, where the Coulomb force between two charged oscillators is linearized for small displacements, introducing an effective  coupling term, $\lambda q_1 q_2$, into the system's Hamiltonian. The strength of this coupling is highly sensitive to experimental parameters such as charge, separation, and frequencies, and with realistic nanofabricated components can attain substantial values relative to the mechanical frequency, facilitating robust macroscopic entanglement\cite{mekonnen2023}. The present work advances beyond these studies by introducing and analyzing a novel hybrid architecture that uniquely integrates such a Coulomb-coupled pair into a triangular ring cavity with a third mechanical oscillator and an intracavity OPA. While prior research has explored Coulomb coupling or OPA enhancement separately in simpler configurations, our system is the first to investigate their combined synergy for the explicit purpose of generating, amplifying, and controlling both bipartite and tripartite entanglement. This integrated approach enables a previously unexplored parameter regime where the Coulomb interaction serves as the primary entanglement generator, while the OPA acts as a powerful, independent control mechanism for dramatic enhancement and shaping of quantum correlations, offering new pathways for multipartite entanglement engineering not accessible in earlier proposals.

In this paper, we study a hybrid electro-optomechan-\\ical system that consists of an optomechanical ring cavity with two charged mechanical oscillators and an uncharged one, containing an OPA. We study how the Coulomb interaction and the OPA can increase bipartite and tripartite entanglement and give an in-depth analysis of the dynamic characteristics and parameter dependence. To this end, we derive the Hamiltonian of the system to obtain the quantum Langevin equations and linearize them around steady-state solutions to account for the Gaussian nature of the quantum fluctuations.

Using the covariance matrix formalism, we quantify entanglement using logarithmic negativity for bipartite subsystems such as between the mechanical modes or between an optical and mechanical mode, and minimum residual contangle for tripartite entanglement.
We analyze how the entanglement can be strengthened depending on the parametric gain, optical phase, detuning, Coulomb coupling strength and laser power.  
\section{Model and dynamics}
\subsection{System description}
\begin{figure}[ht]
	\centering
	\includegraphics[width=0.9\linewidth]{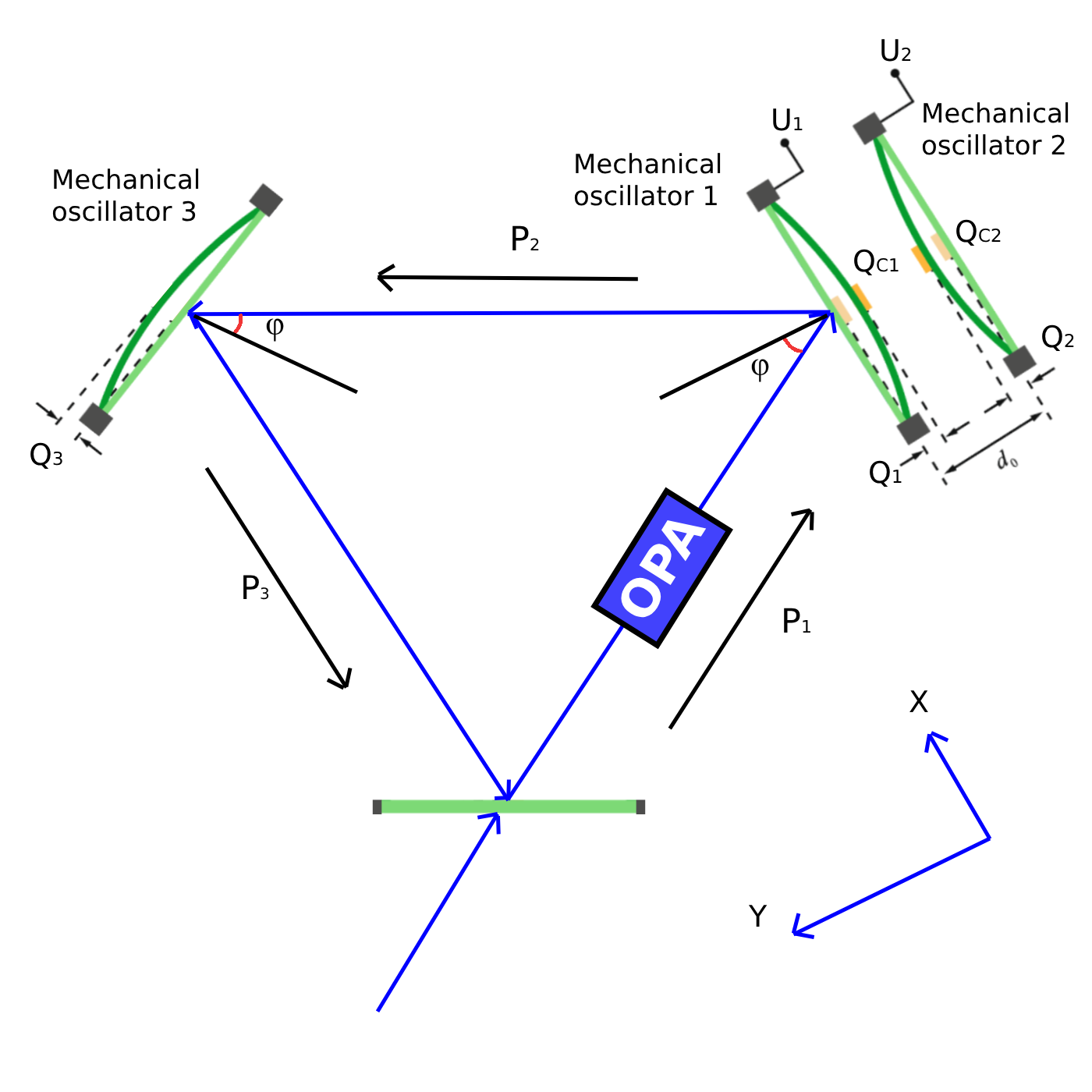}
	\caption{Schematic representation of the optomechanical system}
	\label{fig1}
\end{figure}

This hybrid system consists of an optomechanical ring cavity shaped as a triangle. Inside, a circulating light field interacts via radiation pressure with two perfectly reflective mechanical oscillators. The cavity is formed by a partially transmitting stationary mirror, two charged mechanical oscillators $M_1$ with $Q_{C1}$ and $M_2$ with $Q_{C2}$, coupled via a Coulomb force, and a third mechanical oscillator ($M_3$). These oscillators behave as quantum harmonic oscillators with masses $\mathbf{m}_1$, $\mathbf{m}_2$, and $\mathbf{m}_3$ and resonant frequencies $\omega_{m1}$, $\omega_{m2}$, and $\omega_{m3}$. An OPA inside the cavity, pumped by a secondary laser at frequency $2\omega_L$, creates squeezed light to change the dynamics of the optical field. The cavity is powered by a primary laser operating at frequency $\omega_L$, as shown in Figure \ref{fig1}.
We use the following geometry to derive the optomechanical interaction Hamiltonian $H_{OM}$:
$P_1=(P\sin(\phi),-P\cos(\phi))$, $P_2=(P\sin(\phi),P\cos(\phi))$ and $P_3=(-P,0)$, where $\phi$ is the angle between the incident light and the normal to the mirrors.
\subsection{Hamiltonian}
To accurately describe the system's behavior, the fundamental interactions that control the system's dynamics need to be examined. The Hamiltonian formulation offers an exhaustive mathematical structure that explains the coupling mechanisms between optical and mechanical modes and their subsequent evolution in different scenarios. The Hamiltonian that includes parametric amplification effects allows us to accurately examine the OPA's impact on quantum correlations which facilitates advanced analysis of entanglement enhancement techniques.

The Hamiltonian that describes the system is given by:
\begin{eqnarray}
	\frac{H}{\hbar} = H_0 + H_{\textit{int}} + H_{\textit{drive}} + H_{\textit{OPA}},
\end{eqnarray}
where
\begin{eqnarray}
	H_0 &=& \omega_{c} a^{\dagger}a 
	+ \sum_{i=1}^{3} \frac{\omega_{m_{i}}}{2}(q_{i}^{2} + p_{i}^{2}), \\
	H_{\textit{int}} &=&H_{OM} +H_C, \\
	H_{\textit{drive}} &=& i E_L (a^{\dagger} e^{-i\omega_{L}t} - ae^{i\omega_{L}t}), \\
	H_{\textit{OPA}} &=& i G \left(e^{i \theta} (a^{\dagger})^{2}e^{-2i\omega_{L}t} - e^{- i \theta} a^{2}e^{2i\omega_{L}t}\right).
\end{eqnarray}
The Hamiltonian is composed of several key components. $H_0$ describes the free energy of the cavity optical field and the three mechanical oscillators.

The interaction Hamiltonian, $H_{\textit{int}}$, contains multiple terms: The first term represents radiation pressure interactions between the cavity mode and mechanical oscillators $M_1$ and $M_3$. The radiation pressure force on the first movable mirror arises from the momentum transfer of intracavity photons. Due to the oblique incidence at angle $\phi$, the number of photons striking the mirror during one cavity round-trip time $\Delta t = L/c$ is $n_c \cos\phi$, where $n_c = a^\dagger a$ is the total intracavity photon number. The momentum transferred per photon is $\Delta P = 2\hbar k \cos\phi$, where $k = \omega_c/c$. The resulting radiation pressure force is then given by:
$
F_1 = \left(n_c \cos\phi\right) \times \frac{\Delta P}{\Delta t} = n_c \times \frac{2\hbar\omega_c}{L} \cos^2\phi.
$
The corresponding interaction Hamiltonian is $H_1 = F_1 Q_1$, where $Q_1$ is the mirror's displacement. Expressing $Q_1$ in terms of the dimensionless displacement $q_1$ via $Q_1 = \sqrt{\frac{\hbar}{m_1\omega_{m1}}}q_1$, we obtain:
$
H_1 = \frac{2\hbar\omega_c}{L} \sqrt{\frac{\hbar}{m_1\omega_{m1}}} \cos^2\phi\ a^\dagger a\ q_1.
$
From this, we identify the single-photon optomechanical coupling at normal incidence as:
$
g_1 = \frac{2\omega_c}{L} \sqrt{\frac{\hbar}{m_1\omega_{m1}}}.
$
Following the same process for the oscillating mirror $M_3$ yields the Hamiltonian $
H_3 = -\frac{2\hbar\omega_c}{L} \sqrt{\frac{\hbar}{m_3\omega_{m3}}} \cos^2\phi\ a^\dagger a\ q_3
$, where $g_3 = \frac{2\omega_c}{L} \sqrt{\frac{\hbar}{m_3\omega_{m3}}}$. Finally the optomechanical interaction Hamiltonian is defined as:
\begin{eqnarray}
	H_{OM}= \cos^2(\phi)(g_1 q_{1} - g_3 q_{3}) a^{\dagger} a.
\end{eqnarray}
The second term, $H_C$, is the Coulomb interaction between the charged oscillators $M_1$ and $M_2$. For small displacements ($|Q_1 - Q_2| \ll d_0$), it can be expanded as:

\begin{equation}
	\begin{split}
		H_C&=- \frac{Q_{C1}Q_{C2}}{4\pi\epsilon_0|d_0+Q_1-Q_2|},\\
		&\approx-\frac{Q_{C1} Q_{C2}}{4\pi\epsilon_0d_0} \left[1 - \frac{Q_1 - Q_2}{d_0} + \frac{(Q_1 - Q_2)^2}{d_0^2}\right].
	\end{split}
\end{equation}
After omitting constant energy offsets and eliminating linear terms by shifting to new equilibria, the essential quantum coupling comes from the quadratic cross-term. Expressed in dimensionless quadratures $q_j \equiv Q_j\sqrt{m\omega_m/\hbar}$, this yields $H_C = \lambda q_1 q_2$, with a tunable coupling strength:
\begin{eqnarray}
	\lambda = \frac{Q_{C1}Q_{C2}}{2\pi\epsilon_0 m\omega_m d_0^3}.
\end{eqnarray}
Here, the charges $Q_{Cj} = C_j U_j$ are controlled by gate voltages. 
For typical experimental parameters $ m = 5$~ng, $\omega_m = 2\pi\times10^7$~Hz, $d_0 = 160~\mu$m, $C_1 = C_2 = 12$~pF, $U_1 = U_2 = 100$~V, the value of the Coulomb coupling strength is $\lambda \approx0.32\omega_m$.

Additional terms $H_{\textit{drive}}$ and $H_{\textit{OPA}}$ account for the external laser driving and the OPA, respectively.

Transforming to a frame rotating at the laser frequency $\omega_L$ via $U=e^{i\omega_La^{\dagger}at}$ simplifies the Hamiltonian to:

\begin{align}
	\frac{H_{\textit{Rot}}}{\hbar} 
	&= \Delta_{0} a^{\dagger}a 
	+ \sum_{i=1}^{3} \frac{\omega_{m_{i}}}{2}(q_{i}^{2} + p_{i}^{2})\nonumber\\
	&+ \cos^2(\phi)(g_1 q_{1} - g_3 q_{3}) a^{\dagger} a+ \lambda q_1q_2\nonumber\\
	&+ i E_L (a^{\dagger} - a) + i G \left(e^{i \theta} (a^{\dagger})^{2} - e^{- i \theta} a^{2}\right), \label{H_rot}
\end{align}
where $\Delta_{0} = \omega_{c} - \omega_{L}$ is the cavity detuning. The operators $a$ ($a^{\dagger}$) annihilate (create) a cavity photon, while $q_i$ and $p_i$ are the dimensionless position and momentum operators for the $i$th mechanical oscillator, satisfying $[q_{i}, p_{j}] = i \delta_{ij}$($i=1,2,3$). The single-photon optomechanical coupling for oscillator $k(k=1,3)$ is $g_k = (2\omega_{c}/L) \sqrt{\hbar/(m_k \omega_{m_{k}})}$, the driving amplitude is $E_L = \sqrt{2 \kappa P_L/(\hbar \omega_{L})}$, and the OPA is characterized by its nonlinear gain $G$ and phase $\theta$.

\subsection{Dynamics of the system}
To describe the complete evolution of the system in time, we derive the quantum Langevin equations (QLEs) describing the interaction between optical and mechanical modes, incorporating dissipation and noise. These equations provide a full account of the time evolution of quantum fluctuations. The inclusion of damping and external noise sources ensures our model reflects realistic experimental conditions.

\subsubsection{Derivation of the Nonlinear Quantum Langevin Equations}~\\
The Heisenberg equation of motion for an operator $O$, incorporating damping and noise, is given by:
\begin{equation}
	\dot{O} = -\frac{i}{\hbar}[O, H_{\text{rot}}] + \hat{\mathcal{D}}[O],
\end{equation}
where $\hat{\mathcal{D}}[O]$ represents the Lindblad dissipator. We apply this equation systematically to each system operator using the rotating-frame Hamiltonian $H_{\text{rot}}$ from Eq. \ref{H_rot}. To ensure algebraic accuracy and manage the complexity of the commutation algebra, these steps were implemented and verified using Python's SymPy library for symbolic computation.

\paragraph{Mechanical Quadratures:} For the mechanical position $q_1$, using $[q_1, p_1^2] = 2i p_1$ and $[q_1, q_2] = 0$, we find:
\begin{equation}
	\dot{q}_1 = -\frac{i}{\hbar}[q_1, H_{\text{rot}}] = \omega_{m_1} p_1.
\end{equation}
For the mechanical momentum $p_1$, using $[p_1, q_1^2] = -2i q_1$, $[p_1, a^\dagger a] = 0$, and $[p_1, q_1 q_2] = -i q_2$, and adding the damping ($-\gamma_{m1}p_1$) and noise ($\xi_1$) terms, we obtain:
\begin{equation}
	\dot{p}_1 = -\omega_{m_1} q_1 - \gamma_{m1} p_1 -\lambda q_2 - g_{1}\cos^2(\phi) a^{\dagger}a + \xi_{1}.
\end{equation}

The equations for $q_2, p_2, q_3, p_3$ are derived analogously, employing the canonical commutation relations $[q_j, p_k] = i \delta_{jk}$.

\paragraph{Cavity Mode:} For the cavity annihilation operator $a$, the calculation involves several key commutation relations: $[a, a^\dagger a] = a$, $[a, (a^\dagger)^2] = 2a^\dagger$, and $[a, q_j a] = q_j$ (since $[a, q_j]=0$). Applying these to the Hamiltonian terms yields:

\begin{align}
	-\frac{i}{\hbar}[a, \Delta_0 a^\dagger a] &= -i\Delta_0 a, \\
	-\frac{i}{\hbar}[a, g_1 \cos^2\phi\, q_1 a^\dagger a] &= -i g_1 \cos^2\phi\, q_1 a, \\
	-\frac{i}{\hbar}[a, i G (e^{i\theta}(a^\dagger)^2 - e^{-i\theta}a^2)] &= 2G e^{i\theta} a^{\dagger}.
\end{align}

Adding the cavity damping ($-\kappa a$) and input noise ($\sqrt{2\kappa} a_{\text{in}}$) terms gives the final equation for $\dot{a}$.

Collecting all terms, we arrive at the complete set of nonlinear quantum Langevin equations:
\begin{equation}
	\begin{aligned}
		\dot{q}_{1} &= \omega_{m_{1}} p_{1}, \\
		\dot{p}_{1} &= -\omega_{m_{1}} q_{1} - \gamma_{m1} p_{1} -\lambda q_2-g_{1}\cos^2(\phi) a^{\dagger}a + \xi_{1}, \\
		\dot{q}_{2} &= \omega_{m_{2}} p_{2}, \\
		\dot{p}_{2} &= -\lambda q_1-\omega_{m_{2}} q_{2} - \gamma_{m2} p_{2} + \xi_{2}, \\
		\dot{q}_{3} &= \omega_{m_{3}} p_{3}, \\
		\dot{p}_{3} &= -\omega_{m_{3}} q_{3} - \gamma_{m3} p_{3} +g_{3}\cos^2(\phi) a^{\dagger}a + \xi_{3}, \\
		\dot{a} &= -(\kappa + i \Delta_{0})a - ig_{1} \cos^2(\phi) q_{1}a + ig_{3} \cos^2(\phi) q_{3}a \\
		&+ 2G e^{i\theta} a^{\dagger} + E_L + \sqrt{2 \kappa}\, a_{\text{in}}.
	\end{aligned}
	\label{eq:nonlinear_QLEs}
\end{equation}
Here, $\gamma_{m_{i}}$ ($i = 1,2,3$) and $\kappa$ are the damping rates. The operators $\xi_{i}$ represent Brownian noise sources for the mechanical modes, with zero mean and correlation functions given by \cite{benguria1981quantum}:
\begin{equation}
	\frac{1}{2} \langle \xi_{i}(t) \xi_{i}(t') + \xi_{i}(t') \xi_{i}(t) \rangle \simeq \gamma_{m_{i}} (2 \eta_{\text{th},i} + 1)\delta(t - t'),
\end{equation}
where $\eta_{\text{th},i} = \left[\exp\left(\hbar \omega_{m_{i}}/k_{B} T\right) - 1\right]^{-1}$. The optical input vacuum noise $a_{\text{in}}$ has the correlation \cite{walls2012quantum}:
\begin{equation}
	\langle a_{\text{in}}(t) a_{\text{in}}^\dagger(t') \rangle = \delta(t-t').
\end{equation}

\subsubsection{Steady-State Analysis}~\\
The steady-state values $\alpha$, $q_i^s$, and $p_i^s$ are found by setting all time derivatives and noise terms to zero ($\langle \dot{O} \rangle = 0$) in the nonlinear QLEs, replacing operators by their mean values. This leads to the algebraic system:

\begin{align}
	p_1^s &= 0, \quad p_2^s = 0, \quad p_3^s = 0, \nonumber \\
	q_1^s &= \frac{-(g_{1}\cos^2\phi |\alpha|^2 + \lambda q_{2}^s)}{\omega_{m_{1}}}, \quad 
	q_2^s = \frac{-\lambda q_{1}^s}{\omega_{m_{2}}}, \nonumber \\
	q_3^s &= \frac{g_{3}\cos^2\phi |\alpha|^2}{\omega_{m_{3}}}, \\
	\alpha &= \frac{E_L}{\kappa - 2G \cos\theta + i(\Delta - 2G \sin\theta)},\nonumber
\end{align}

where the effective detuning is $\Delta = \Delta_{0} + (g_1 q_{1}^s - g_3 q_{3}^s)\cos^2\phi$.

\subsubsection{Linearization Procedure and Linearized Equations}~\\
We linearize the dynamics by expanding each operator around its steady-state value: $q_i = q_i^s + \delta q_i$, $p_i = p_i^s + \delta p_i$, $a = \alpha + \delta a$. Substituting these into the nonlinear QLEs and neglecting second-order fluctuation terms (e.g., $\delta a^\dagger \delta a$, $\delta q_i \delta a$) constitutes the linearization approximation. This is valid in the regime of a strong coherent drive ($|\alpha| \gg 1$), where fluctuation amplitudes are small relative to the coherent field. The symbolic computation was instrumental in performing this expansion systematically, isolating the linear terms, and verifying the cancellation of zeroth-order terms.

We then introduce the optical quadrature fluctuations:

\begin{align*}
	\delta x &= (\delta a + \delta a^{\dagger})/\sqrt{2}, &
	\delta y &= (\delta a - \delta a^{\dagger})/(i\sqrt{2}), \\
	\delta x_{\text{in}} &= (\delta a_{\text{in}} + \delta a_{\text{in}}^{\dagger})/\sqrt{2}, &
	\delta y_{\text{in}} &= (\delta a_{\text{in}} - \delta a_{\text{in}}^{\dagger})/(i\sqrt{2}),
\end{align*}

and define the linearized optomechanical coupling strengths $G_1 = \sqrt{2}g_{1}\cos^2\phi\, \alpha$ and $G_3 = \sqrt{2}g_{3}\cos^2\phi\, \alpha$ (taking $\alpha$ to be real without loss of generality). This yields the set of linearized QLEs:

\begin{equation}
	\begin{aligned}
		\delta \dot{q}_{1} &= \omega_{m_{1}} \delta p_{1}, \\
		\delta \dot{p}_{1} &= -\omega_{m_{1}} \delta q_{1} - \gamma_{m1} \delta p_{1} -\lambda \delta q_2 - G_1 \delta x + \xi_{1}, \\
		\delta \dot{q}_{2} &= \omega_{m_{2}} \delta p_{2}, \\
		\delta \dot{p}_{2} &= -\lambda \delta q_1 - \omega_{m_{2}} \delta q_{2} - \gamma_{m2} \delta p_{2}, \\
		\delta \dot{q}_{3} &= \omega_{m_{3}} \delta p_{3}, \\
		\delta \dot{p}_{3} &= - \omega_{m_{3}} \delta q_{3} - \gamma_{m3} \delta p_{3} + G_3 \delta x + \xi_{3}, \\
		\delta \dot{x} &= (-\kappa + 2G \cos\theta) \delta x + (\Delta + 2G \sin\theta) \delta y + \sqrt{2 \kappa}\, \delta x_{\text{in}}, \\
		\delta \dot{y} &= -G_1 \delta q_{1} + G_3 \delta q_{3} - (\Delta - 2G\sin\theta) \delta x \\
		&-(\kappa + 2G\cos\theta) \delta y + \sqrt{2 \kappa}\, \delta y_{\text{in}}.
	\end{aligned}
	\label{eq20}
\end{equation}

\subsubsection{Matrix Formulation and Drift Matrix}~\\
The linearized QLEs can be expressed compactly as:
\begin{equation}
	\dot{u}(t) = A u(t) + N(t),
	\label{eq:linear_matrix_form}
\end{equation}
where $u(t) = (\delta q_{1}, \delta p_{1}, \delta q_{2}, \delta p_{2}, \delta q_{3}, \delta p_{3}, \delta x, \delta y)^{T}$ is the vector of fluctuation operators, and $N(t)$ represent the noise vector, where:\\ $N(t) = (0, \xi_{1}, 0, \xi_{2}, 0, \xi_{3}, \sqrt{2 \kappa}\, \delta x_{\text{in}}, \sqrt{2 \kappa}\, \delta y_{\text{in}})^{T}$. The $8 \times 8$ drift matrix $A$ encapsulates the system's deterministic dynamics. Its elements, derived from the coefficients of the linearized QLEs, are given by:

\begin{equation}
	A = 
	\left(\begin{smallmatrix}
		0 & \omega_{m_1} & 0 & 0 & 0 & 0 & 0 & 0 \\
		-\omega_{m_1} & -\gamma_{m_1} & -\lambda & 0 & 0 & 0 & -G_1 & 0 \\
		0 & 0 & 0 & \omega_{m_2} & 0 & 0 & 0 & 0 \\
		-\lambda & 0 & -\omega_{m_2} & -\gamma_{m_2} & 0 & 0 & 0 & 0 \\
		0 & 0 & 0 & 0 & 0 & \omega_{m_3} & 0 & 0 \\
		0 & 0 & 0 & 0 & -\omega_{m_3} & -\gamma_{m_3} & G_3 & 0 \\
		0 & 0 & 0 & 0 & 0 & 0 & -\kappa + 2G \cos\theta & \Delta + 2G \sin\theta \\
		-G_1 & 0 & 0 & 0 & G_3 & 0 & -(\Delta - 2G \sin\theta) & -(\kappa + 2G \cos\theta)
	\end{smallmatrix}\right).
\end{equation}

\subsection{Covariance matrix}
A steady state is reached when all eigenvalues of the drift matrix $A$ have negative real parts \cite{dejesus1987routh}. Given the linearized dynamics and Gaussian noise, the quantum fluctuations are completely described by an $8 \times 8$ covariance matrix $V$. This matrix is the solution to the Lyapunov equation \cite{vitali20072}:

\begin{equation}
	A V + V A^{T} = -D.
\end{equation}

Here, $D$ is the noise correlation matrix, defined via
\begin{equation}
	\left\langle N_i(t) N_j(t') + N_j(t') N_i(t)\right\rangle / 2 = D_{ij} \delta (t - t').
\end{equation}
For our model, this matrix is diagonal:
\begin{equation}
	\begin{split}
		& D = \mathrm{diag}\bigl(0,\ \gamma_{m_{1}}(2 n_{th_{1}} + 1),\ 0,\ \gamma_{m_{2}}(2 n_{th_{2}} + 1),\ 0,\ \\
		&\gamma_{m_{3}}(2 n_{th_{3}} + 1),\ \kappa,\ \kappa\bigr).
	\end{split}
\end{equation}
\section{Quantification of quantum correlations}
\subsection{Bipartite entanglement}
Quantum correlations in the bipartite system are assessed via the logarithmic negativity $E_N$ \cite{vidal2002computable,plenio2005logarithmic}, a well-established measure for Gaussian entanglement. For a state described by the covariance matrix $\sigma$, partitioned into blocks $X$, $Y$ (local modes) and $Z$ (correlations),
\begin{equation}
	\sigma = \left(
	\begin{array}{cc}
		X & Z \\
		Z^{T} & Y \\
	\end{array}
	\right),
\end{equation}
the logarithmic negativity is given by:
 \begin{equation}
 	E_N=\max[0,-\ln(2\varrho)].
 \end{equation}
 The value $\varrho$ is the minimum symplectic eigenvalue of the partially transposed state, defined as:
\begin{equation}
	\varrho\equiv2^{-1/2}\left\{\Psi-\left[\Psi^{2}-4\det(\sigma)\right]^{1/2}\right\}^{1/2},
	\label{EN}
\end{equation}
with $\Psi=\det(X)+\det(Y)-2\det(Z).$
A non-zero value of $E_N$ ($\varrho < 1/2$) is a direct signature of entanglement, fulfilling the necessary and sufficient conditions of Simon's criterion \cite{simon2000}.

\subsection{Tripartite entanglement}
To quantify the tripartite entanglement, we use the minimum residual contangle \cite{adesso2006}, $\mathcal{R}^{\textrm{min}}_{\tau}$, which is a \textit{bona fide} CV often defined simply as:

\begin{equation}
	\mathcal{R}^{\textrm{min}}_{\tau} = \min_{(r,s,t)} \big[E^{r|st}_{\tau} - E^{r|s}_{\tau} - E^{r|t}_{\tau} \big],
\end{equation}

where $(r,s,t)$ represents all permutations of the three\\-mode indices. Here, $E_{\tau}^{u|v}$ is the contangle of the subsys-\\tem $u$ (one mode) with subsystem $v$ (one or two modes), measured with a suitable entanglement monotone, here the square of the logarithmic negativity.

The one-mode-versus-one-mode contangles $E^{r|s}_{\tau}$ and $E^{r|t}_{\tau}$ can directly be obtained from their definition $E_{\tau} \equiv [E_{\mathcal{N}}]^2$. The computation of the one-mode-versus-two-modes contangle $E^{r|st}_{\tau}$ is related to a nontrivial adjustment of the usual definition appearing in Eq. \ref{EN}. More precisely, the formula for $\varrho$ needs to be rewritten as:

\begin{align}  
	\varrho_{r|st} = \min \big[ \text{eig} |\big[\bigoplus_{k=1}^{3} (-\sigma_{y})\big] P_{r|st} V P_{r|st}| \big],  
\end{align} 

where $\varrho$ is the smallest symplectic eigenvalue of the partial transposition of the $6 \times 6$ covariance matrix $V$ with $\sigma_{y}$ the Pauli matrix in the $y$- direction and $P_{r|st} = \text{diag}(1,-1,1,1,1,1)$ is the corresponding partial transposition matrix. On the other hand, the partial transposition matrices for the other bipartitions read $P_{s|rt} = \text{diag}(1,1,1,-1,1,1)$ and $P_{t|sr} =\text{diag}(1,1,1,1,1,-1)$. Moreover, the residual contangle must hold the condition according to the Coffman-Kundu-Wootters monoga-\linebreak my inequality for quantum entanglement $E^{r|st}_{\tau} - E^{r|s}_{\tau} - E^{r|t}_{\tau} \geq 0.$

\section{Numerical Implementation and Stability Analysis}
The entanglement dynamics were computed through a systematic numerical procedure implemented in Python. For each parameter set, we first calculated the steady-state optical amplitude $\alpha$ and the linearized coupling strengths $G_{1,3} = \sqrt{2}g_{1,3}|\alpha|$. The core of the calculation involved constructing the drift matrix $A$ and the noise correlation matrix $D$ from their analytical forms. Crucially, we performed a linear stability analysis at each point: the system was deemed stable only if all eigenvalues of $A$ had negative real parts ($\Re(\lambda_i) < 0$). For stable configurations, we solved the continuous-time Lyapunov equation $A V + V A^T = -D$ for the covariance matrix $V$ using SciPy's \texttt{solve\_lyapunov} function. From the $4\times4$ submatrix of $V$ corresponding to the chosen bipartition (e.g., modes $M_3$ and $C$), we extracted the symplectic eigenvalues to compute the logarithmic negativity $E_N = \max[0, -\ln(2\varrho)]$, where $\varrho$ is the minimum symplectic eigenvalue of the partially transposed state. This automated pipeline, combining stability checks with covariance matrix analysis, generated the curves and similar results throughout the paper.

\section{Results and discussion}
We now present a numerical analysis of the quantum correlations in the electro-optomechanical ring cavity system, focusing on how key parameters influence both bipartite and tripartite entanglement. The entanglement is quantified using the logarithmic negativity ($E_N$) for bipartite subsystems and the minimum residual contangle ($\mathcal{R}^{\min}_{\tau}$) for tripartite correlations. All simulations are performed using these accessible experimental parameters: a laser frequency of $\omega_L = 2\pi \times 3.7 \times 10^{14}$ Hz, mechanical frequencies $\omega_{m1} = \omega_{m2} = \omega_{m3} = \omega_m = 2\pi \times 10^7$ Hz, mechanical damping rates $\gamma_m = 2\pi \times 10^2$ Hz, a cavity decay rate $\kappa = \pi \times 10^7$ Hz, and optomechanical coupling strengths $g_1 = g_3 = 2\pi \times 35$ Hz.

Our investigation begins by examining the role of the OPA in enhancing bipartite entanglement, particularly between the two Coulomb-coupled mechanical modes, $M_1$ and $M_2$. We demonstrate that adjusting the OPA gain $G$ relative to the cavity decay rate $\kappa$ induces a significant increase in entanglement. This underscores optical parametric amplification as a critical mechanism for strengthening quantum correlations. Subsequently, we analyze the effects of other parameters including optical phase ($\theta$), Coulomb coupling strength ($\lambda$), laser detuning ($\Delta$), and input power ($P_L$) on both the stability and the magnitude of the generated entanglement.

\begin{figure*}[!t] 
	\centering
	\includegraphics[width=1\linewidth]{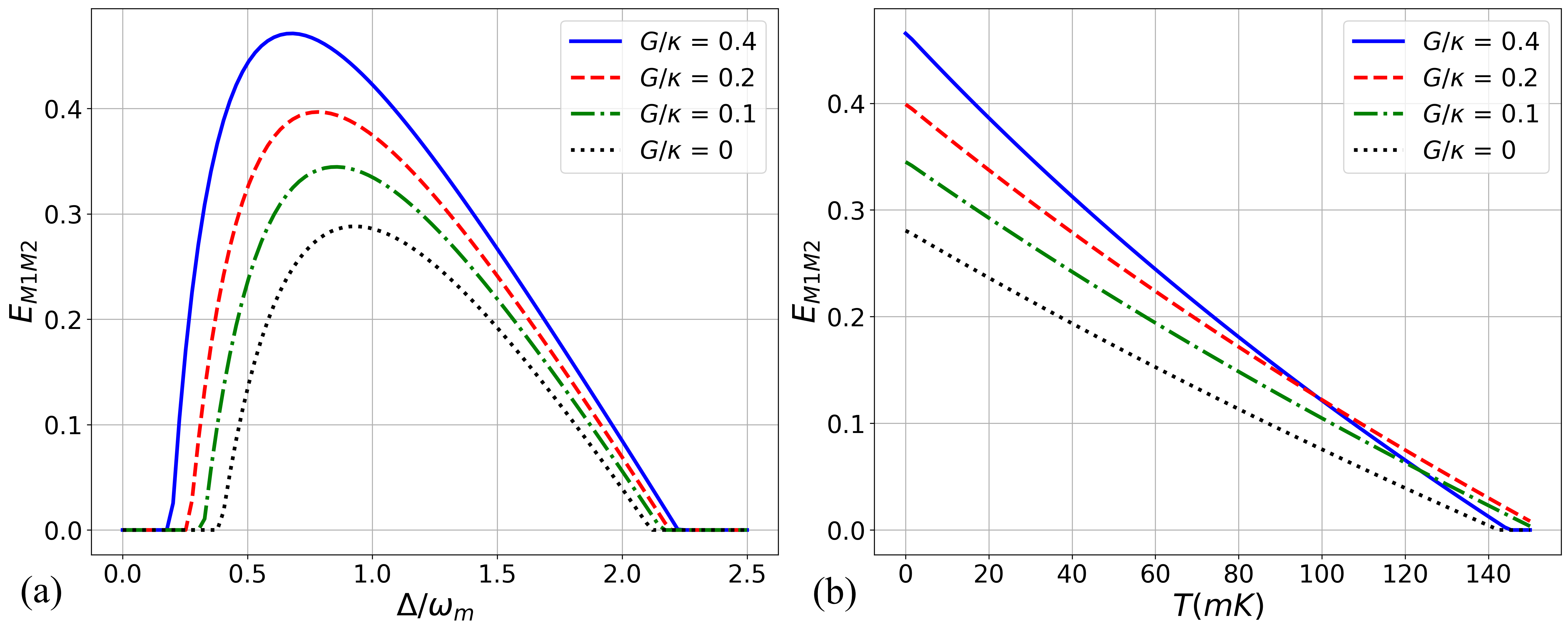}
	\caption{(a) Plot of logarithmic negativity $E_{M1M2}$ versus the normalized detuning $\Delta/\omega_m$. We have taken $\theta=\pi$,  $\lambda=0.9\omega_m$, $P_L=60mW$ and $T=1mK$ . (b) $E_{M1M2}$ versus temperature for a fixed value of normalized detuning $\Delta=0.8\omega_m$. The remaining parameters are the same as in Figure 2(a).
	}
\label{fig2}
\end{figure*}

The first key result is presented in Fig.~\ref{fig2}(a), which shows the variation of logarithmic negativity $E_{M1M2}$ quantifying the entanglement between the two mechanical modes without any contribution of the OPA ($G=0$), and with the existence of the OPA, as a function of the normalized detuning $\Delta/\omega_m$. In the absence of the OPA ($G / \kappa = 0$), the maximum logarithmic negativity of about 0.287 is obtained at a normalized detuning of approximately 0.912. At $G / \kappa = 0.1$, the resulting maximal logarithmic negativity increases to $\sim 0.344$ around $\Delta / \omega_m \approx 0.844$. For $G / \kappa = 0.2$, the maximum is pushed to approximately 0.397 at $\Delta / \omega_m \approx 0.784$. For $G / \kappa = 0.4$, the logarithmic negativity peak reaches approximately 0.471 at $\Delta / \omega_m \approx 0.674$. This indicates that an appropriate choice of the OPA gain could enhance the maximum entanglement and broaden the range of detuning over which appreciable entanglement persists. We attribute the results to the critical importance of optical parametric amplification in amplifying quantum correlations.

In Fig. \ref{fig2} (b), we plot the logarithmic negativity $E_{M1M2}$ as a function of temperature for several values of the gain ratios $G/\kappa$. As expected, we notice that as the temperature $T$ increases, $E_{M1M2}$ decreases until it reaches zero value. The degradation of entanglement with increasing temperature follows from the thermal decoherence of the mechanical modes. As thermal fluctuations increase, they introduce noise that disrupts quantum correlations, eventually leading to complete entanglement suppression.

\begin{figure*}[!t] 
	\centering
	\includegraphics[width=1\linewidth]{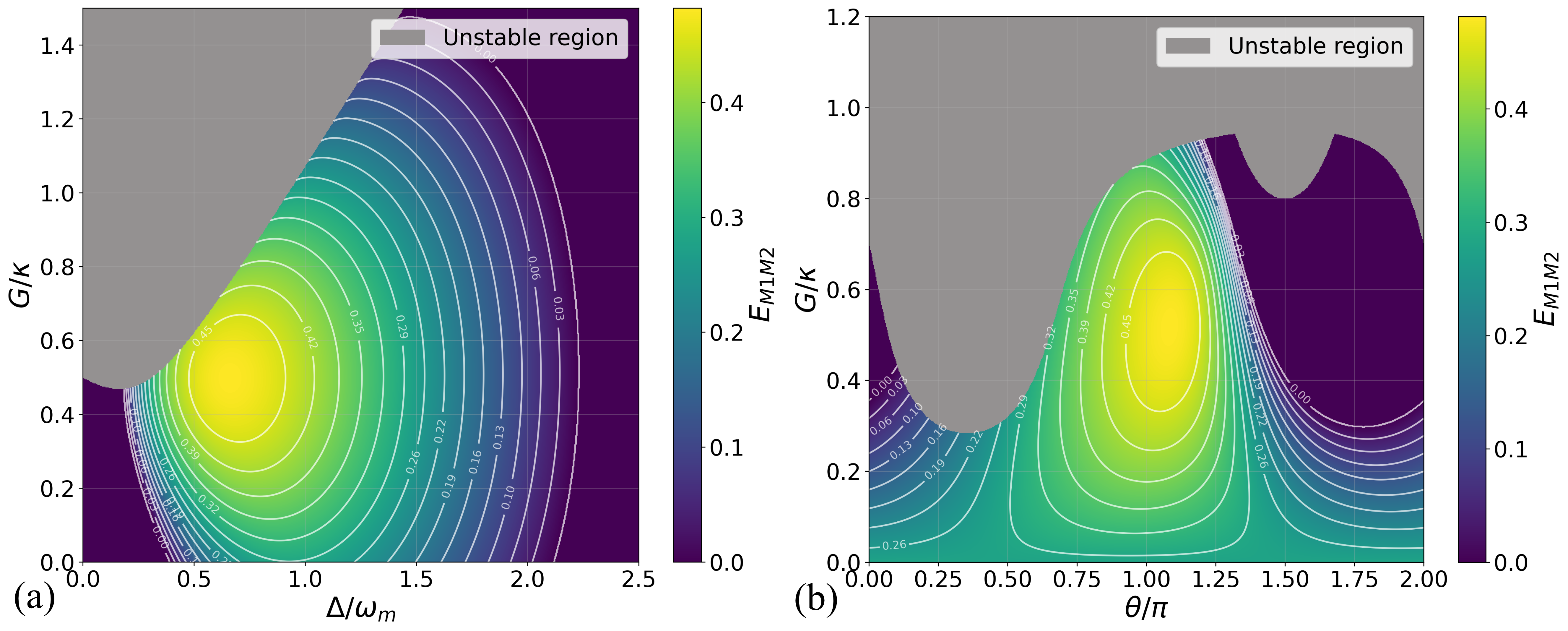}
	\caption{
		(a) $E_{M1M2}$ as function of the $\Delta/\omega_m$ and $G/\kappa$. The chosen parameters are: $P_L=60mW$, $T=1mK$, $\lambda=0.9\omega_m$ and $\theta=\pi$. (b) $E_{M1M2}$ as function of the phase ratio $\theta/\pi$ and $G/\kappa$ for $\Delta=0.8\omega_m$. The other parameters are the same as in Figure 3(a).
	}
	\label{fig3}
\end{figure*}

Figure~\ref{fig3}(a) shows the region where the system becomes unstable. We can see that for $G/\kappa > 0.5$, the allowed values of detuning are reduced with the increase of the OPA gain. This occurs because the amplification process of the OPA injects additional energy into the system, which must be balanced by choosing the appropriate parameters such as the laser detuning $\Delta$, the OPA phase $\theta$ to avoid runaway oscillations and maintain stability. While higher gain can enhance quantum entanglement, it simultaneously makes the system more susceptible to instability, requiring more precise control over parameters.
To further investigate the dependence of entanglement on the optical phase $\theta$, we plot the logarithmic negativity $E_{M1M2}$ as a function of the phase ratio $\theta/\pi$ and the gain ratio $G / \kappa$ in Fig.~\ref{fig3}(b). In the absence of the OPA ($G / \kappa = 0$), the logarithmic negativity remains constant at a value of approximately 0.28. As the gain ratio increases, the entanglement is enhanced within the phase region \( 0.5 \leq \theta/\pi \leq 1.3 \), while it decreases elsewhere. The maximum logarithmic negativity of approximately 0.48 is achieved at $\theta \approx \pi$ and $G/\kappa \approx 0.5$. These results demonstrate that both the OPA gain $G$ and the phase $\theta$ are essential for entanglement optimization. To enhance entanglement, an optimal phase $\theta$ must be selected, as specific phases correspond to maximal entanglement while others diminish it. This dependence on the optical phase $\theta$ underscores the importance of coherent control in quantum systems. By tuning $\theta$, one can modulate the interference effects between different interaction pathways, thereby controlling whether quantum correlations are enhanced or reduced through constructive or destructive interference.

\begin{figure*}[!t]
	\centering
	\includegraphics[width=1\linewidth]{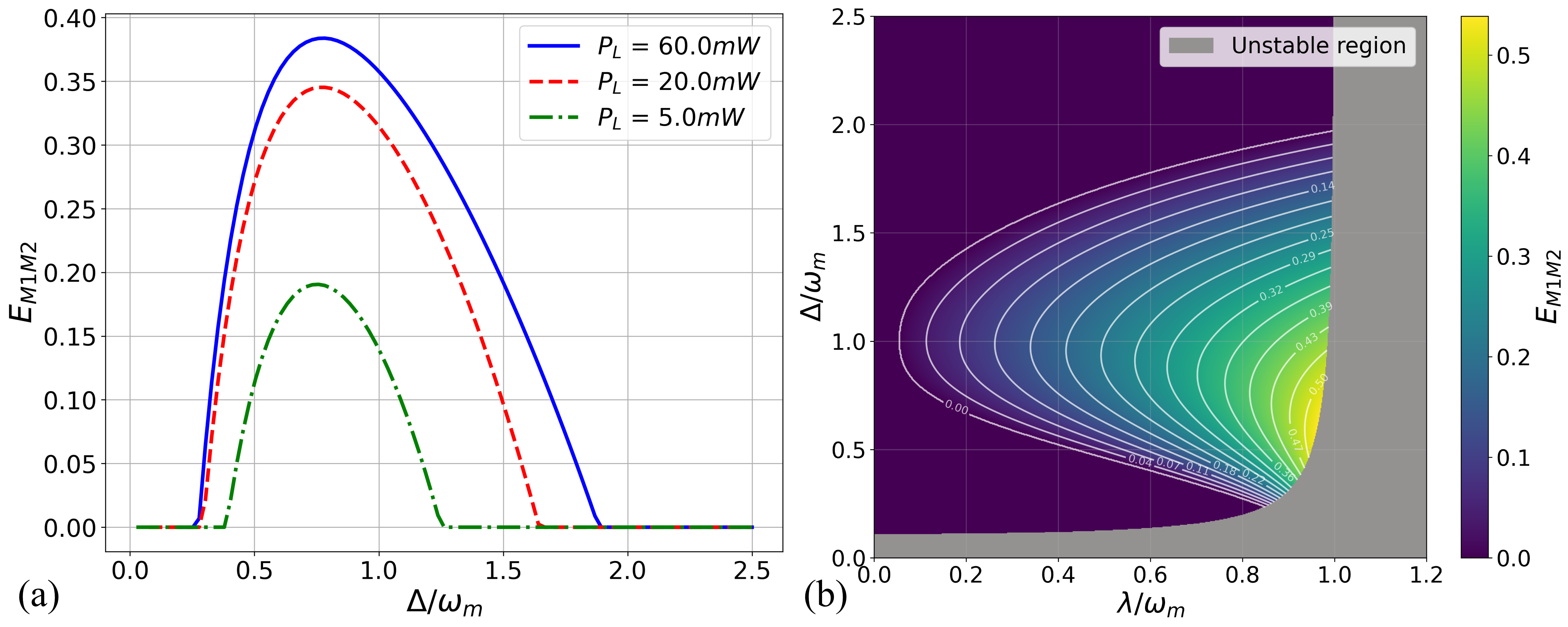}
	\caption{(a) $E_{M1M2}$ vs $\Delta/\omega_m$ for various values of input LASER, the other parameters are: $G=0.5\kappa$, $\theta=\pi$,  $\lambda=0.8\omega_m$ and $T=5mK$. (b) $E_{M1M2}$ vs $\Delta/\omega_m$ for different values of the coupling $\lambda$ where $P_L=60mW$. The remaining parameters are the same as in Figure 4(a).}
	\label{fig4}
\end{figure*}

Now we examine the effect of the power of LASER on entanglement, we find as shown in Fig. \ref{fig4}(a) that the entanglement can be improved in strength by raising the incident power of the input laser. The principle behind this improvement of the entanglement of the optomechanical system is the enhancement of radiation pressure force acting on the movable mirror of the optomechanical cavity. This effect of enhancement has, of course, its limits since an increase in input laser power will amplify extra noise inside the system bringing entanglement to destruction.

\begin{figure*}[!t]
	\centering
	\includegraphics[width=1\linewidth]{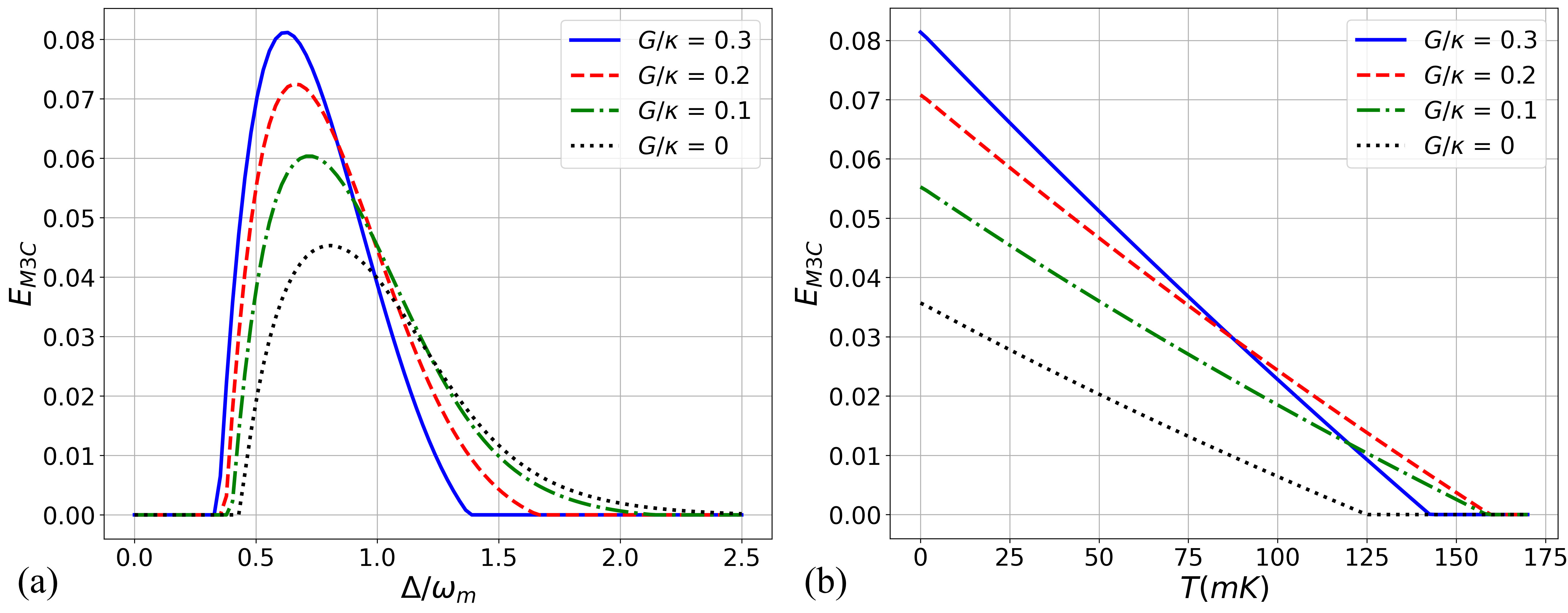}
	\caption{(a) Plot of logarithmic negativity $E_{M3C}$ versus the normalized detuning $\Delta/\omega_m$ for different values of the gain ratios $G/\kappa$.  We have taken $\theta=\frac{\pi}{2}$, $\lambda=0.2\omega_m$ and $P_L=50mW$. (b) $E_{M3C}$ versus temperature for a fixed value of $\Delta/\omega_m=0.6$ }
	\label{fig5}
\end{figure*}
Fig. \ref{fig4}(b) shows the entanglement between the two mechanical modes $E_{M1M2}$ as a function of the normalized detuning $\Delta/\omega_m$ and normalized coupling strength $\lambda/\omega_m$. The results demonstrate a clear dependence of quantum entanglement on the Coulomb coupling strength, $\lambda$. Notably, entanglement is entirely absent in the absence of coupling ($\lambda=0$). As $\lambda$ increases, it drives a more efficient coherent energy exchange, which amplifies the maximum entanglement and extends the detuning range over which these quantum correlations persist. For instance, a maximum logarithmic negativity of approximately 0.52 is achieved for $\lambda=0.95\omega_m$ at a detuning of $\Delta=0.65\omega_m$. However, the system loses its stability as $\lambda$ approaches the mechanical frequency $\omega_m$, indicating a fundamental operational limit.

Fig. \ref{fig5} (a) shows the logarithmic negativity $E_{M3C}$ describing entanglement between the third mechanical mode and the optical mode versus normalized detuning $\Delta / \omega_m$ for various $G / \kappa$ values. For zero gain ($G / \kappa = 0$), the logarithmic negativity is non-significant, maximal value is approximately 0.045 at $\Delta / \omega_m=0.795$. For $G / \kappa= 0.1$, the entanglement is boosted, reaching a peak logarithmic negativity of around 0.06 at $\Delta / \omega_m=0.719$. At higher gain $G / \kappa = 0.3$, logarithmic negativity is at its peak value $0.81$. It is obvious from these results that increasing the gain ratio gives a significant improvement in quantum entanglement. Furthermore, we notice, as shown in Fig. \ref{fig5}(b), that the entanglement decreases as the temperature increases; however, the presence of the OPA helps maintain entanglement by compensating for thermal decoherence, thereby mitigating the detrimental effects of temperature on quantum correlations. 

\begin{figure}[ht]
	\centering
	\includegraphics[width=1\linewidth]{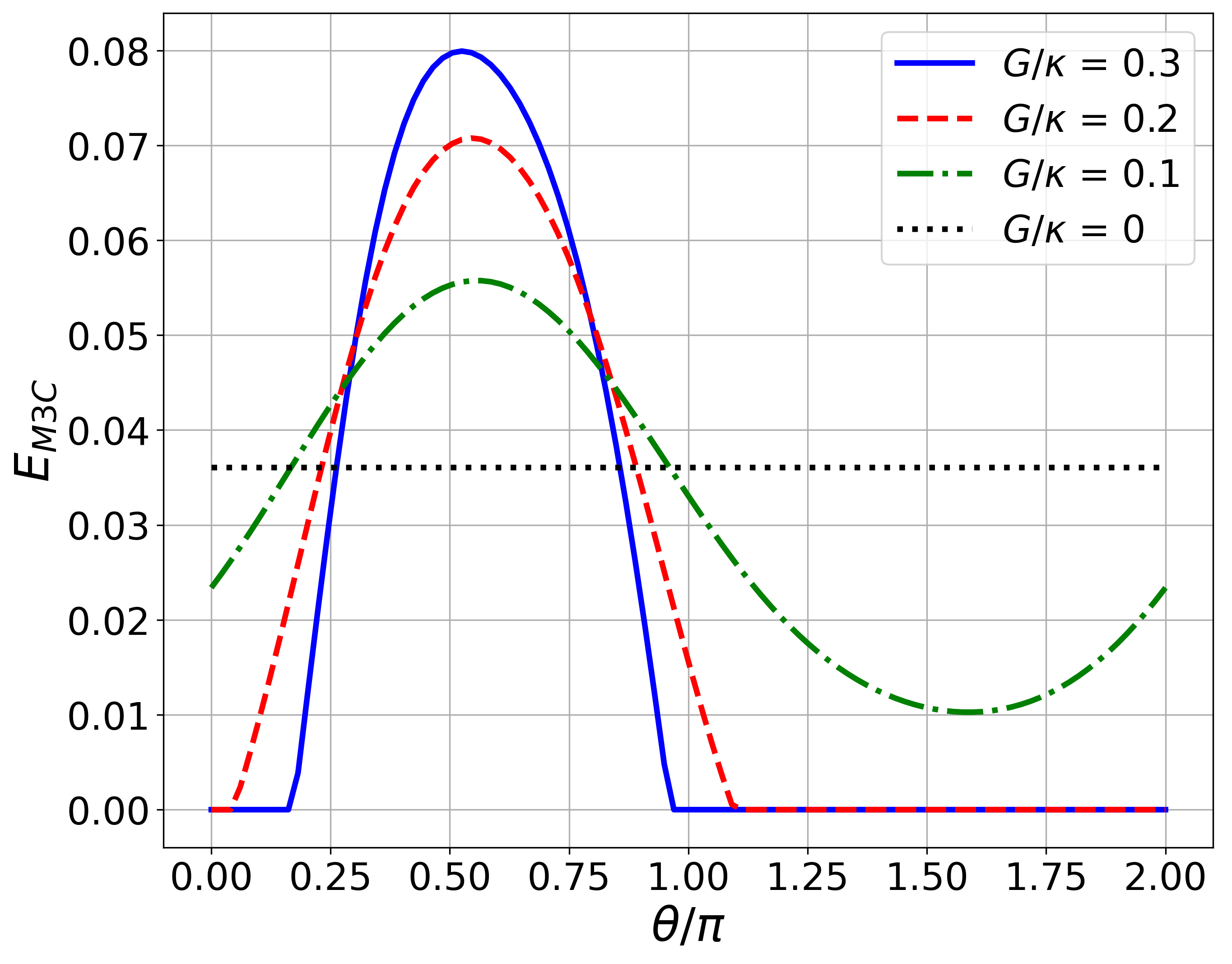}
	\caption{$E_{M3C}$ versus $\theta/\pi$ for different values of $G/\kappa$.  Other parameters are $\Delta/\omega_m=0.6$, $\lambda=0.2\omega_m$,  $P_L=50mW$ and $T=1mK$.}
	\label{fig6}
\end{figure}

In Fig. \ref{fig6}, we plot the logarithmic negativity $E_{M3C}$ as a function of the optical phase ratio $\theta/\pi$ in the absence of the OPA ($G = 0$) and for different values of the gain ratios. For $G=0$, the logarithmic negativity has a constant value of approximately $0.036$. As the gain ratio $G/\kappa$ increases, the maximum entanglement also increases, reaching about $0.056$ for $G/\kappa = 0.1$, $0.071$ for $G/\kappa = 0.2$, and $0.08$ for $G/\kappa = 0.3$. These results show that both the optical phase ratio $\theta/\pi$ and the gain ratio $G/\kappa$ play a crucial role in enhancing entanglement. A proper choice of phase and gain can significantly boost the entanglement, highlighting the importance of optimizing these parameters for improved quantum correlations.

The parameter-dependent entanglement landscapes presented in Figs.~\ref{fig5}(a) and \ref{fig6}(a) are a direct manifestation of the nonlinear control enabled by the OPA. The OPA's gain $G$ and phase $\theta$ are not merely amplitude controls; they are the key parameters of a tunable nonlinearity that fundamentally reshapes the system's interaction Hamiltonian. The enhancement of entanglement with increasing $G$, followed by the saturation and instability at high gain, exemplifies a classic nonlinear system trade-off: the injected parametric energy first amplifies the desired quantum correlations but eventually drives the system towards an unstable regime. This behaviour is conceptually analogous to the management of soliton amplitude and stability in nonlinear wave systems governed by similar parametric terms \cite{liu2025gardner, gao2024ytsf}. Therefore, our results demonstrate that the OPA functions as an essential nonlinear element for entanglement engineering, where its parameters allow precise navigation of the quantum correlation landscape.

\begin{figure}[ht]
	\centering
	\includegraphics[width=1\linewidth]{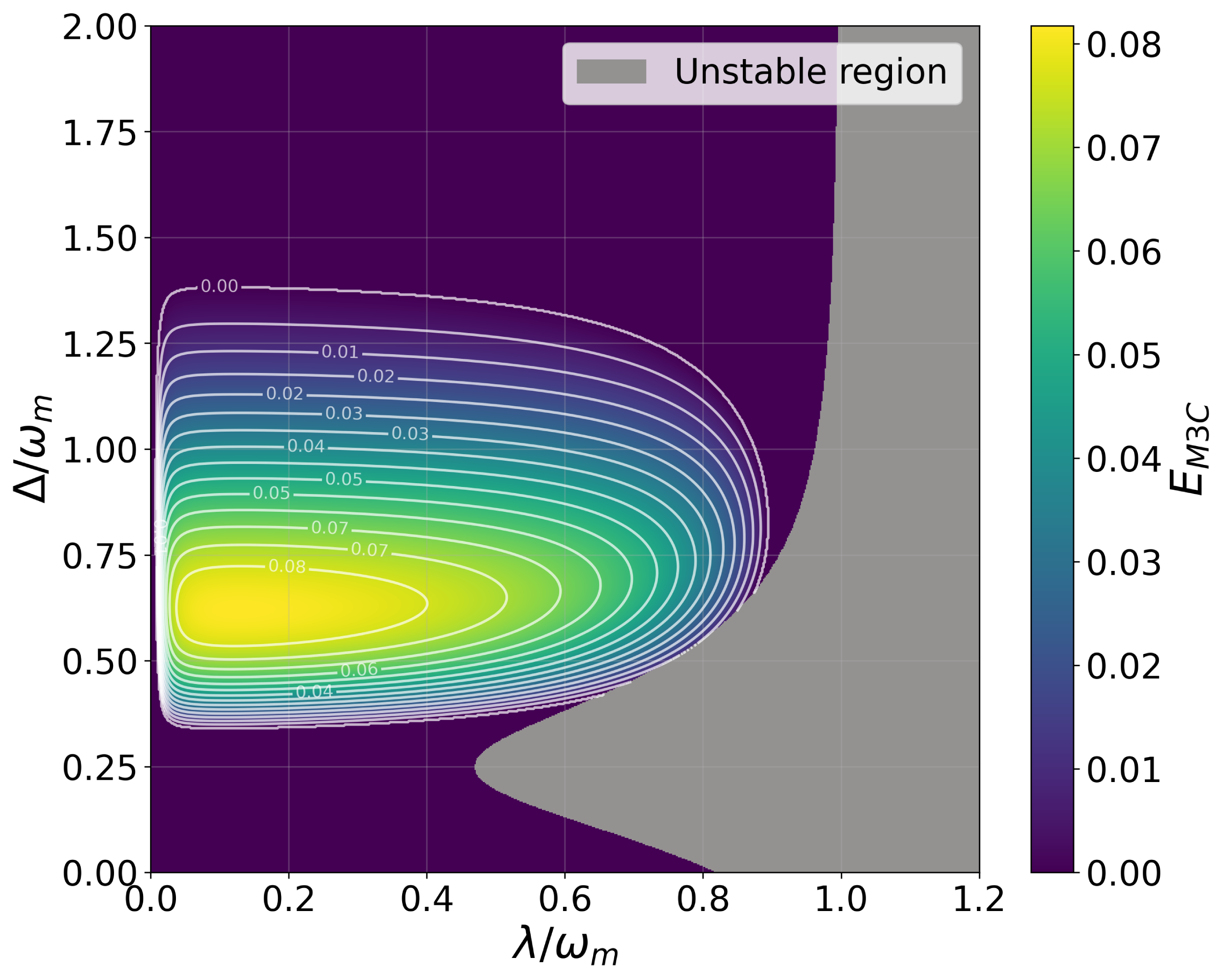}
	\caption{$E_{M3C}$ versus $\Delta/\omega_m$ and $\lambda/\omega_m$. Other parameters are:  $P_L=50mW$, $T=1mK$, $G=0.3\kappa$ and $\theta=\pi/2$.}
	\label{fig7}
\end{figure}

Fig. \ref{fig7} presents the dependence of the bipartite entanglement, quantified by logarithmic negativity $E_{M3C}$, between the third mechanical mode and the optical cavity mode on the Coulomb coupling strength $\lambda/\omega_m$, the most salient result is the non-monotonic relationship between the Coulomb coupling $\lambda$ and the generated entanglement. The entanglement is entirely absent at $\lambda = 0$, confirming that the Coulomb interaction is the indispensable driver for this quantum correlation. A well-defined maximum in $E_{M3C}$ is observed for intermediate coupling strengths in the range of $0.1 < \lambda/\omega_m < 0.3$, with the peak value reaching $E_{M3C} \approx 0.08$. However, for stronger coupling beyond $\lambda/\omega_m > 0.4$, the entanglement monotonically decreases, eventually vanishing entirely at $\lambda \approx 0.9\omega_m$. These results imply that the Coulomb interaction optimizes the system's dynamics, creating conditions that enhance the transfer of entanglement from the optical field to the mechanically isolated mode $M_3$. Consequently, $\lambda$ functions both as a direct coupling mechanism and a global modulator of quantum correlations across the entire hybrid system.

\begin{figure*}[!t]
	\centering
	\includegraphics[width=1\linewidth]{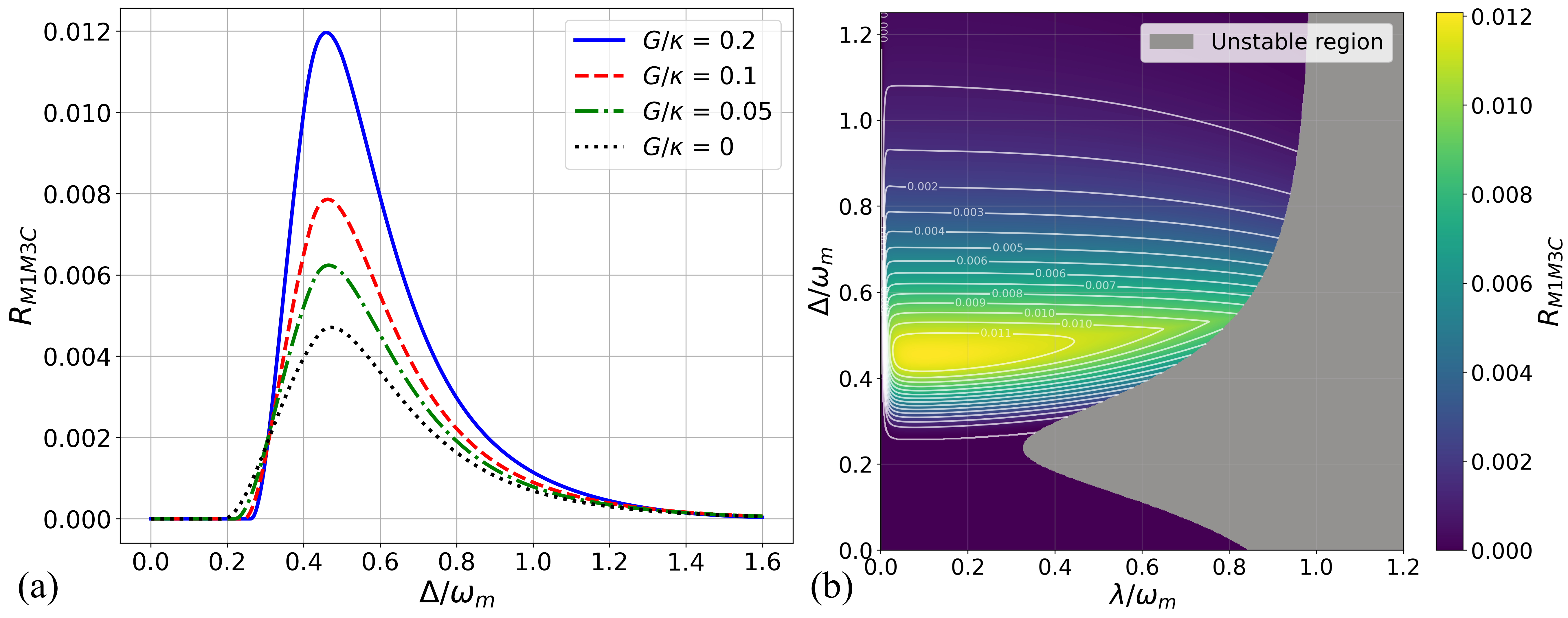}
	\caption{(a) Plot of minimum residual contangle $R_{M1M3C}$ versus the normalized detuning $\Delta/\omega_m$ for different values of the gain ratios $G/\kappa$. We have taken $\theta=0.37\pi$, $\lambda=0.2\omega_m$,  $P_L=60mW$ and $T=1mK$. (b) $R_{M1M3C}$ versus $\Delta/\omega_m$ and $\lambda/\omega_m$ for $G=0.2\kappa$. The remaining parameters are the same as in Figure 8(a). }
	\label{fig8}
\end{figure*}
\begin{figure*}[!t]
	\centering
	\includegraphics[width=1\linewidth]{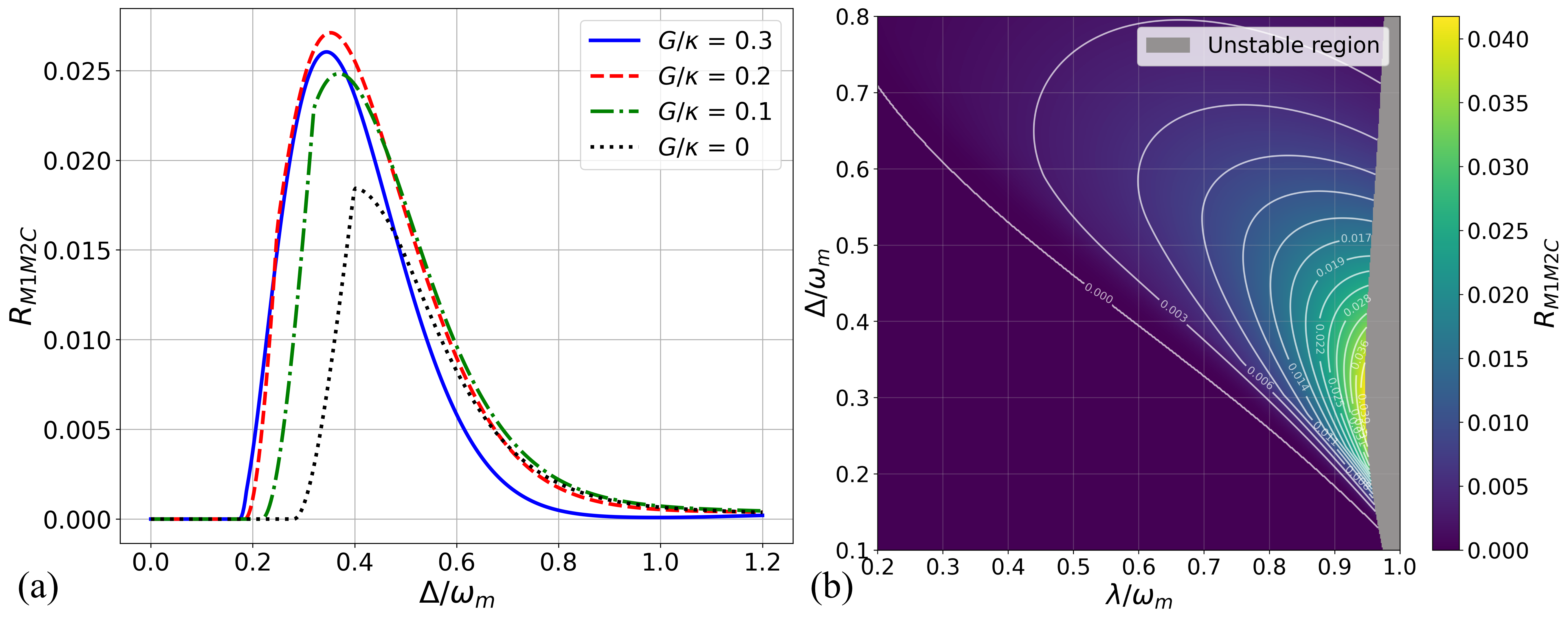}
	\caption{(a) Plot of minimum residual contangle $R_{M1M2C}$ versus the normalized detuning $\Delta/\omega_m$ for different values of the gain ratios $G/\kappa$. We have taken $\theta=\pi$, $\lambda=0.9\omega_m$,  $P_L=60mW$ and $T=1mK$. (b) $R_{M1M3C}$ versus $\Delta/\omega_m$ and $\lambda/\omega_m$ for $G=0.2\kappa$. The remaining parameters are the same as in Figure 9(a). }
	\label{fig9}
\end{figure*}
We show, in Fig. \ref{fig8}(a), the variation of the minimum residual contangle $R_{M1M3C}$ that quantifies tripartite entanglement between the first, third and the optical modes, as a function of the normalized detuning $\Delta / \omega_m$, for different gain ratio $G / \kappa$. We find that in the absence of any OPA inside the ring cavity ($G = 0$), the maximum residual contangle is $R_{\min} \approx 0.0045$, occurring at $\Delta / \omega_m \approx 0.48$. As the gain ratio increases, entanglement is enhanced.
For $G / \kappa = 0.2$, the maximum value of $R_{\min}$ reaches approximately $0.012$ at the same value of $\Delta / \omega_m$.
These results show the crucial role of the OPA in enhancing tripartite entanglement. With a suitable choice of gain ratio and optical phase $\theta$, entanglement can be optimized. Furthermore, the ring cavity arrangement provides stable and strong entanglement due to the cyclic energy transfer between the optical and mechanical modes enhancing the optomechanical interaction. This setup plus a parametric gain gives a powerful control to manage quantum correlations on optomechanical systems.
Figure \ref{fig8}(b) shows the variation of $R_{M1M3C}$ as a function of $\Delta/\omega_m$ and $\lambda/\omega_m$. As expected, entanglement is absent when $\lambda=0$. The plot reveals that the highest entanglement value, approximately 0.012, occurs at $\Delta/\omega_m \approx 0.45$ for lower values of the Coulomb coupling strength $\lambda$.

Figure \ref{fig9}(a) depicts the tripartite entanglement between the first mechanical, second mechanical, and optical modes for various ratios of $G/\kappa$. In the absence of an OPA, the entanglement reaches a maximum of only 0.018 at $\Delta = 0.4\omega_m$. However, as the gain $G$ increases, the entanglement is significantly enhanced. We observe a maximum entanglement of approximately 0.027 at $\Delta = 0.35\omega_m$, indicating that the peak shifts toward lower detunings with increasing $G$.
The effect of Coulomb coupling on the tripartite entanglement dynamics is presented in Fig.~\ref{fig9}(b). A threshold coupling strength of $\lambda \approx 0.43\omega_m$ is required to generate entanglement between the first mechanical, second mechanical, and optical modes. The maximum entanglement of $0.04$ is achieved at an optimal parameter set of $\lambda = 0.94\omega_m$ and $\Delta = 0.31\omega_m$. This analysis underscores the importance of parameter optimization (namely $\lambda$, $\Delta$, $G$, and $\theta$) for enhancing quantum correlations. A key finding is the trade-off associated with strong coupling: although higher $\lambda$ values yield greater entanglement, they also increase the system's propensity for instability, imposing a requirement for more exacting parameter precision.

Our findings can be formally stated as consequences of two mathematical results: (1) For the linearized Gaussian system, steady-state entanglement exists if and only if the Lyapunov equation $A V + V A^T = -D$ yields a covariance matrix $V$ whose appropriate submatrices violate the Positive Partial Transpose (PPT) criterion for separability. This violation is quantified by the minimum symplectic eigenvalue $\varrho$ of the partially transposed state, with $\varrho < 1/2$ being a necessary and sufficient condition for bipartite entanglement. (2) The stability thresholds observed in parameter scans (e.g., $G/\kappa > 0.5$) are exact solutions to the Routh-Hurwitz conditions applied to the characteristic polynomial of the drift matrix $A$. Thus, each numerical curve of $E_N(G, \theta, \lambda, \Delta)$ or $\mathcal{R}^{\min}_{\tau}(G, \theta, \lambda, \Delta)$ represents a particular realization of these general mathematical conditions under the specific nonlinear parametric drive of the OPA.

\section{Conclusion}
In this work, we have presented a comprehensive theoretical analysis of quantum correlation generation in a hybrid electro-optomechanical system. The setup, consisting of a ring cavity with two Coulomb-coupled mechanical resonators and an embedded OPA, provides a rich platform for controlling both bipartite and tripartite entanglement.

Our study yields several key insights. First, we have unequivocally demonstrated that the Coulomb interaction is an indispensable driver for establishing entanglement, not only directly between the charged mechanical modes but across the entire system. Its influence mediates and enhances quantum correlations even between the optical cavity mode and the mechanically isolated oscillator, acting as a global modulator of the system's quantum dynamics. Second, and crucially, the OPA serves as a powerful tool for the significant further enhancement of these quantum correlations. By tuning the OPA's gain and phase, we can dramatically amplify the degree of entanglement for all tested bipartite subsystems as well as genuine tripartite entanglement.

Furthermore, our exploration of the parameter space revealed that the robustness and magnitude of entanglement are highly sensitive to the laser detuning, input power, and Coulomb coupling strength, and can be optimized by appropriately tuning these parameters. However, this enhancement is accompanied by a critical trade-off: operating at high gain or strong coupling to maximize entanglement simultaneously constrains the stable operating regime of the system, necessitating precise parameter control.  Moreover, we confirmed that thermal noise degrades entanglement, emphasizing the need for low-temperature operation. These results highlight the crucial role of parameter selection in maximizing entanglement and provide insights for designing robust optomechanical platforms for quantum technologies.




\begin{thebibliography}{99} 

\bibitem{Jammer1966}
M. Jammer, 
\textit{The Conceptual Development of Quantum Mechanics.} (McGraw-Hill, New York, 1966)

\bibitem{Pais1982} 
A. Pais, 
\textit{Subtle is the Lord: The Science and the Life of Albert Einstein} (Oxford University Press, Oxford, 1982)

\bibitem{Aspect2002} 
A. Aspect, 
\textit{Bell's Theorem: The Naive View of an Experimentalist}, in \textit{Quantum [Un]speakables: From Bell to Quantum Information} (Springer, Berlin, 2002)

\bibitem{einstein1935can}
A. Einstein, B. Podolsky and N. Rosen, 
\textit{Phys. Rev.} \textbf{47}, 777 (1935)

\bibitem{Bell1987} 
J. S. Bell, 
\textit{Speakable and Unspeakable in Quantum Mechanics} (Cambridge University Press, Cambridge, 1987).

\bibitem{bell1964on}
J. S. Bell, 
\textit{Physics Physique Fizika} \textbf{1}, 195 (1964)

\bibitem{aspect1981experimental}
A. Aspect, P. Grangier and G. Roger, 
\textit{Phys. Rev. Lett.} \textbf{47}, 460 (1981)

\bibitem{aspect1982experimental}
A. Aspect, P. Grangier and G. Roger, 
\textit{Phys. Rev. Lett.} \textbf{49}, 91 (1982)

\bibitem{aspect1982experimental2}
A. Aspect, J. Dalibard and G. Roger, 
\textit{Phys. Rev. Lett.} \textbf{49}, 1804 (1982)

\bibitem{Bennett1984} 
C. H. Bennett and G. Brassard, \textit{Quantum Cryptography: Public Key Distribution and Coin Tossing} (Bangalore, 1984)

\bibitem{Nielsen2010}
M. A. Ni\-els\-en \-and I. L. Ch\-ua\-ng,
\textit{Quant\-um Computa\-tion and Quan\-tum Inform\-ation} (Camb\-ridge Univ\-ersity Pr\-ess, Camb\-ridge, 2010)

\bibitem{kimble2008quantum}
H. J. Kimble, 
\textit{Nature} \textbf{453}, 1023 (2008)

\bibitem{ekert1991quantum}
A. K. Ekert, 
\textit{Phys. Rev. Lett.} \textbf{67}, 661 (1991)

\bibitem{pathak2013elements}
A. Pathak, 
\textit{Elements of Quantum Computation and Quantum Communication} (CRC Press, Boca Raton, 2013)

\bibitem{aspelmeyer2014}
M. Aspelmeyer, T. J. Kippenberg and F. Marquardt, 
\textit{Rev. Mod. Phys.} \textbf{86}, 1391 (2014)

\bibitem{kippenberg2008}
T. J. Kippenberg and K. J. Vahala, 
\textit{Science} \textbf{321}, 1172 (2008)

\bibitem{Bowen2015} 
W. P. Bowen and G. J. Milburn, 
\textit{Quantum Optomechanics} (CRC Press, Boca Raton, 2015)

\bibitem{caves1980}
C. M. Caves, 
\textit{Phys. Rev. Lett.} \textbf{45}, 75 (1980)

\bibitem{Xia2023}
Xia, Y., Agrawal, A.R., Pluchar, C.M. et al,
\textit{Nat. Photon.} \textbf{17}, 470–477 (2023)

\bibitem{simon2011}
R. Ghobadi, A. R. Bahrampour and C. Simon, 
\textit{Phys. Rev. A} \textbf{84}, 033846 (2011)

\bibitem{youssefi2023}
Youssefi, A., Kono, S., Chegnizadeh, M. et al.
\textit{Nat. Phys.} \textbf{19}, 1697–1702 (2023)

\bibitem{chan2011}
J. Chan et al., 
\textit{Nature} \textbf{478}, 89 (2011)

\bibitem{vitali2007}
D. Vitali et al., 
\textit{Phys. Rev. Lett.} \textbf{98}, 030405 (2007)

\bibitem{amazioug2018}
M. Amazioug, M. Nassik and N. Habiballah, 
\textit{Int. J. Quantum Inf.} \textbf{16}, 1850043 (2018)

\bibitem{xu2022optomechanically}
H. Xu, Y. Ma and X. Zhang, 
\textit{Nat. Phys.} \textbf{18}, 172 (2022)

\bibitem{li2021cavity}
J. Li, S. Gr\"{o}blacher and M. Poot, 
\textit{Adv. Quantum Technol.} \textbf{4}, 2100030 (2021)

\bibitem{brawley2016nonlinear}
G. A. Brawley, M. R. Vanner, P. E. Larsen, S. Schmid, A. Boisen and W. P. Bowen, 
\textit{Nat. Commun.} \textbf{7}, 10988 (2016)

\bibitem{kippenberg2007cavity}
T. J. Kippenberg and K. J. Vahala, 
\textit{Science} \textbf{321}, 1172 (2008)

\bibitem{marquardt2011quantum}
F. Marquardt and S. M. Girvin, 
\textit{Physics} \textbf{2}, 40 (2009)

\bibitem{huang2009entangling}
S. Huang and G. S. Agarwal, 
\textit{New J. Phys.} \textbf{11}, 103044 (2009)

\bibitem{schulze2010optomechanical}
R. J. Schulze and C. Genes, 
\textit{Phys. Rev. A} \textbf{81}, 063810 (2010)

\bibitem{huang2009}
S. Huang and G. S. Agarwal, 
\textit{Phys. Rev. A} \textbf{79}, 013821 (2009)

\bibitem{agarwal2016}
G. S. Agarwal and S. Huang, 
\textit{Phys. Rev. A} \textbf{93}, 043844 (2016)

\bibitem{mi2013}
X. Mi, J. Bai and S. Ke-hui, 
\textit{Eur. Phys. J. D} \textbf{67}, 109 (2013)

\bibitem{hu2017}
C. S. Hu, X. R. Huang, L. T. Shen, Z. B. Yang and H. Z. Wu, 
\textit{Eur. Phys. J. D} \textbf{71}, 109 (2017)

\bibitem{xuereb2012}
A. Xuereb, M. Barbieri and M. Paternostro, 
\textit{Phys. Rev. A} \textbf{86}, 013809 (2012)

\bibitem{gao2025kmm}
X.-Y. Gao, J.-G. Liu, G.-W. Wang,
\textit{Appl. Math. Lett.} \textbf{171}, 109615 (2025)

\bibitem{gao2025sk}
X.-Y. Gao,
\textit{Appl. Math. Lett.} \textbf{159}, 109262 (2025)

\bibitem{gao2025hs}
X.-Y. Gao,
\textit{China Ocean Eng.} \textbf{39}, 541--547 (2025)

\bibitem{gao2025zkb}
X.-Y. Gao, X.-Q. Chen, Y.-J. Guo, W.-R. Shan,
\textit{Qual. Theory Dyn. Syst.} \textbf{24}, 73 (2025)

\bibitem{gao2024water}
X.-Y. Gao,
\textit{Chin. J. Phys.} \textbf{92}, 1233--1239 (2024)

\bibitem{shan2024kdv}
H.-W. Shan, B. Tian, C.-D. Cheng, X.-T. Gao, Y.-Q. Chen, H.-D. Liu,
\textit{Qual. Theory Dyn. Syst.} \textbf{23} (Suppl. 1), 267 (2024)

\bibitem{feng2025fluid}
C.-H. Feng, B. Tian, X.-T. Gao,
\textit{Qual. Theory Dyn. Syst.} \textbf{24}, 100 (2025)

\bibitem{wang2026hd}
G.-W. Wang, Z.-X. Tan, X.-Y. Gao, J.-G. Liu,
\textit{Appl. Math. Lett.} \textbf{172}, 109720 (2026)

\bibitem{gao2025kvcbs}
X.-Y. Gao, W.-G. Zhao, D.-W. Zuo,
\textit{Qual. Theory Dyn. Syst.} \textbf{24}, 234 (2025)

\bibitem{zhang2012}
J. Q. Zhang, Y. Li, M. Feng and Y. Xu, 
\textit{Phys. Rev. A} \textbf{86}, 053806 (2012)

\bibitem{bai2017}
C. H. Bai, D. Y. Wang, H. F. Wang, A. D. Zhu and S. Zhang, 
\textit{Sci. Rep.} \textbf{7}, 2545 (2017)

\bibitem{mekonnen2023}
Mekonnen, H.D., Tesfahannes, T.G., Darge, T.Y. et al, 
\textit{Sci. Rep.} \textbf{13}, 13800 (2023)

\bibitem{benguria1981quantum}
R. Benguria and M. Kac, 
\textit{Phys. Rev. Lett.} \textbf{46}, 1 (1981)

\bibitem{walls2012quantum}
D. F. Walls and J. D. Harvey, 
\textit{Quantum Optics VI: Proceedings of the Sixth International Symposium on Quantum Optics} (Springer, 2012)

\bibitem{dejesus1987routh}
E. X. DeJesus and C. Kaufman, 
\textit{Phys. Rev. A} \textbf{35}, 5288 (1987)

\bibitem{vitali20072}
D. Vitali, S. Mancini and P. Tombesi, 
\textit{J. Phys. A: Math. Theor.} \textbf{40}, 8055 (2007)

\bibitem{vidal2002computable}
G. Vidal and R. F. Werner, 
\textit{Phys. Rev. A} \textbf{65}, 032314 (2002)

\bibitem{plenio2005logarithmic}
M. B. Plenio, 
\textit{Phys. Rev. Lett.} \textbf{95}, 090503 (2005)

\bibitem{simon2000}
R. Simon, 
\textit{Phys. Rev. Lett.} \textbf{84}, 2726 (2000)

\bibitem{adesso2006}
G. Adesso and F. Illuminati, 
\textit{New J. Phys.} \textbf{8}, 15 (2006)


\bibitem{liu2025gardner}
H.-D. Liu, B. Tian, Y.-Q. Chen, T.-Y. Zhou, X.-T. Gao,
\textit{Nonlinear Dyn.} \textbf{113}, 3655--3672 (2025)

\bibitem{gao2024ytsf}
X.-Y. Gao,
\textit{Appl. Math. Lett.} \textbf{152}, 109018 (2024)

\end{thebibliography}
\end{document}